\documentclass[11pt]{amsart}

\usepackage[square,compress,comma, numbers,sort]{natbib}
\usepackage[colorlinks=true, citecolor=red, linkcolor=blue]{hyperref}
\usepackage{amsfonts,mathtools}

\allowdisplaybreaks[4]

\usepackage{amssymb}
\usepackage{color}
\newcommand{\abs}[1]{\left\lvert #1 \right\rvert}

\DeclarePairedDelimiterXPP\pk[1]{\mathbb{P}}\{ \}{}{ #1}
\DeclarePairedDelimiterXPP\E[1]{\mathbb{E}}\{ \}{}{	#1}

 \usepackage{xparse}

\NewDocumentCommand{\ceil}{s O{} m}{%
  \IfBooleanTF{#1} 
    {\left\lceil#3\right\rceil} 
    {#2\lceil#3#2\rceil} 
}
\NewDocumentCommand{\floor}{s O{} m}{%
  \IfBooleanTF{#1} 
    {\left\lfloor#3\right\rfloor}
    {#2\lfloor#3#2\rfloor}
}

\def\x{\vk{x}}

\newtheorem{theo}{Theorem}[section]
\newtheorem{sat}[theo]{Proposition}
\newtheorem{de}[theo]{Definition}
\newtheorem{lem}[theo]{Lemma}

\newtheorem{example}[theo]{Example}
\newtheorem{korr}[theo]{Corollary}
\newtheorem{remark}[theo]{Remark}

\numberwithin{equation}{section}

\newcommand{\prooftheo}[1]{ \textsc{Proof of Theorem} \ref{#1} }
\newcommand{\proofprop}[1]{\textsc{Proof of Proposition} \ref{#1}}
\newcommand{\prooflem}[1]{\textsc{Proof of Lemma} \ref{#1}}

\newcommand{\QED}{\hfill $\Box$}

\newcommand{\COM}[1]{}

\def\IF{\infty}

\newcommand{\R}{\mathbb{R}}
\newcommand{\inr}{\in \R}

\newcommand{\BQN}{\begin{eqnarray}}
\newcommand{\EQN}{\end{eqnarray}}
\newcommand{\BQNY}{\begin{eqnarray*}}
	\newcommand{\EQNY}{\end{eqnarray*}}
\def\ldot{, \ldots,}
\def\bar{\overline}

\newcommand{\limit}[1]{\lim_{#1 \to   \infty}}
\def\todis{\overset{d}\rightarrow}

\newcommand{\kb}[1]{\boldsymbol{#1}}
\newcommand{\vk}[1]{\kb{#1}}

\def\bqny#1{\begin{eqnarray*}#1\end{eqnarray*}}
\def\bqn#1{\begin{eqnarray}#1\end{eqnarray}}

\newcommand{\BS}{\begin{sat}}
	\newcommand{\ES}{\end{sat}}
\newcommand{\BT}{\begin{theo}}
	\newcommand{\ET}{\end{theo}}
\newcommand{\BK}{\begin{korr}}
	\newcommand{\EK}{\end{korr}}

\newcommand{\BEX}{\begin{example}}
	\newcommand{\EEX}{\end{example}}

\newcommand{\BD}{\begin{de}}
	\newcommand{\ED}{\end{de}}

\newcommand{\BIT}{\begin{itemize}}
	\newcommand{\EIT}{\end{itemize}}
\newcommand{\BDI}{\begin{description}}
	\newcommand{\EDI}{\end{description}}

\newcommand{\BRM}{\begin{remark}}
	\newcommand{\ERM}{\end{remark}}

\newcommand{\BEL}{\begin{lem}}
	\newcommand{\EEL}{\end{lem}}
\newcommand{\nelem}[1]{{Lemma \ref{#1}}}

\newcommand{\netheo}[1]{{Theorem \ref{#1}}}

\def\X{\vk{X}}

\def\X{\vk{X}}

\newcommand\expon[1]{\exp\left(#1\right)}

\begin{document}

\title[Multivariate Risk Measures for Gaussian Risks]
{\small Approximation of Some Multivariate Risk Measures for Gaussian Risks}

\author{Enkelejd  Hashorva}
\address{Enkelejd Hashorva, Department of Actuarial Science, University of Lausanne,\\
Chamberonne 1015 Lausanne, Switzerland}
\email{Enkelejd.Hashorva@unil.ch}

\bigskip

\date{\today}
 \maketitle

\begin{quote}

{\bf Abstract:} Gaussian random vectors exhibit the loss of dimension phenomena, which relate to their joint survival tail behaviour. Besides, the fact that the components of such vectors are light-tailed complicates the  approximations of various multivariate risk measures significantly. In this contribution we derive precise  approximations of  marginal mean excess, marginal expected shortfall and multivariate conditional tail expectation of Gaussian random vectors and highlight  links with conditional limit theorems. Our study indicates that similar results hold for  elliptical and Gaussian like multivariate risks. 
\end{quote}

 

\section{Introduction}
The recent article \cite{Das1} investigates two important measures of risk contagion for a given bivariate random vector $(Z_1,Z_2)$, namely the   marginal mean   excess  (MME) and the marginal expected shortfall (MES). 
Specifically, under the assumption that $\E{\abs{Z_1}}< \IF$ the MME   is defined  for any $p\in (0,1)$ by 
\bqn{ E(p) = \E{ (Z_1 - VaR_{Z_2}(p))_+ \lvert Z_2> VaR_{Z_2}(p)}, 
 }
whereas MES is given as 
\bqn{ S(p) = \E{ Z_1  \lvert Z_2> VaR_{Z_2}(p)}, 
}
with $VaR_{Z_i}(p)$ the Value-at-Risk at level $p$ of $Z_i$, which is simply the quantile function of $Z_i$ at $p$. In general both 
 $E(p)$ and $S(p)$ cannot be calculated explicitly. Besides, in the risk management practice the main interest is the calculation of these quantities for $p$ being close to  1. \\ 
   In this paper we shall consider first the approximations of MME and MES for $(Z_1,Z_2)$ being jointly Gaussian with  correlation $\rho \in (-1,1)$. Gaussian random vectors are asymptotically independent, i.e., large values occur independently which in our context means that   
  $$ \lim_{p \uparrow  1} \pk{ Z_1 > VaR_{Z_1}(p)  \lvert Z_2 > VaR_{Z_2}(p)  } =0.$$
  Moreover, Gaussian risks  exhibit the dimension reduction phenomenon, i.e.,  the joint survival probability can be proportional to the marginal survival probability for large values of the threshold, see e.g., \cite{ENJH02,Hashorva05,MR2397662} and the discussion below. Indeed that  phenomenon renders the approximations  of both MME and MES interesting and challenging. 
  
Under hidden regular variation assumption on $(Z_1,Z_2)$ the recent publications  \cite{Das1,Das2} consider approximations of MME and MES  under some additional asymptotic conditions.  However the Gaussian setup is not covered therein since the marginal distributions are in our setup light-tailed. As discussed recently in \cite{AsmussenE}, see also \cite{Nolde} the light-tailed case is very challenging (even in the one-dimensional setup) and surprisingly very little investigated in the literature.  \\
  Given the central role of multivariate Gaussian distributions, and the interesting behaviour of light-tailed risks, our principal goal in this contribution is to derive approximations of MME and MES in the Gaussian setup. We state next the result for the bivariate case.
 
Throughout in the following  $\Phi$  denotes the distribution function (df) of an $N(0,1)$ random variable with inverse $\Phi^{-1}$  and $\varphi  $ the probability density function (pdf) of a standard Gaussian random vector $(X_1,X_2)$ with correlation $\rho \in (-1,1)$.

\BT Let $\vk Z=(Z_1, Z_2)$ be jointly Gaussian with $Z_i$ having  $N(\mu_i, \sigma_i^2),i=1,2$   df and correlation $\rho \in (-1,1)$ and set $u_p= \Phi^{-1}(p),  \beta= (\mu_2- \mu_1)/\sigma_1, \eta = \beta /\sqrt{1- \rho^2}$.\\
i) If $ \sigma_2> \rho \sigma_1  $ and  $ \sigma_1> \rho \sigma_2  $, then  
\bqn{ \label{stda}
E(p) &\sim &   \frac{\sigma_1}{ h_1^2h_2} \sqrt{ 2 \pi}  u_p^{-2} e^{ \frac{ u_p^2}{2}} \varphi  ( \sigma_2    u_p/\sigma_1+ \beta ,  u_p) \to 0, \quad p\uparrow  1, }
where  
\BQNY 
  h_1= \frac{\sigma_2 - \rho \sigma_1 }{\sigma_1(1- \rho^2)} >0, \quad h_2=\frac{\sigma_1- \rho \sigma_2 }{\sigma_1(1- \rho^2)}>0.
\EQNY
ii) If $\sigma_2 = \rho \sigma_1  $, then 
\bqn{ \lim_{p\uparrow 1} E(p) &= & \sigma_1 
	 \sqrt{1- \rho^2} \Bigl( \Phi'( \eta ) -  \eta  [1- \Phi( \eta)  ] \Bigl) \in (0, \IF). 
\label{stda2}	}
iii) If $\sigma_2 < \rho \sigma_1$, then 
\bqn{  
	E(p) &\sim &(\rho \sigma_1 - \sigma _2) u_p\to \IF , \quad p\uparrow  1.   
\label{e15}}
iv) If $\sigma_1 \le  \rho \sigma_2  $, then 
\bqn{\label{i6}  E(p) &\sim&   \sigma_1  e^{  -\frac{\beta^2 }{2 }}   \Phi( \eta  \rho^*  ) 
	u_p e^{- \beta \frac{\sigma_2}{\sigma_1} u_p} e^{ -\frac{\sigma_2^2 - \sigma_1^2}{2 \sigma_1^2}u_p^2}\to 0, \quad p\uparrow 1,   
}
where $\rho^*=\rho$ if $\sigma_2= \rho \sigma_1$ and $\rho^*=\IF$ otherwise. \\
v) As $p\uparrow 1$ we have 
\bqn{
	 \label{eqe1} S(p)  - \mu_1-  \sigma_1 \rho u_p &\to& 0.}
\label{th1}
\ET 

The above findings show that $E(p)$ and $S(p)$ have a completely different behaviour as $p$ approaches 1.  Both  \eqref{stda} and \eqref{i6} prove  that $E(p)$ tends super-exponentially fast to 0 as $p\to 1$. A completely different behaviour is observed in \eqref{stda2} and \eqref{e15}. 
For the approximation of MES we have only one case as shown in \eqref{eqe1}, since  its definition is invariant to $\sigma_2$. \\ 
The bivariate setup is however restrictive; it is possible to have in \eqref{eqe1} a non-zero limit in higher dimensions, see Remark \ref{remkoka}. Indeed, the two-dimensional setup is easier to deal with and there are no additional notation needed, but it does not show how to derive corresponding results in multivariate setup.  
 
It is worth mentioning  that extensions of our results to elliptical random vectors 
are also possible, but those require more technical efforts and additional assumptions similar to \cite{Jaworski}[Assumption 4]. Moreover, extensions to the larger class of  Gaussian like random vectors treated in \cite{farkas2017asymptotic} 
 can also be obtained, but again  further technical treatments are needed  and will therefore not be addressed here. Besides, our findings  are of certain importance for considering approximations of other risk measures such as multivariate expectiles considered in \cite{maume2017multivariate}. 

Brief outline of the rest of the paper: In the next section we focus on the multivariate setup deriving the approximations of MME, MES and the multivariate conditional tail expectation (MCTE). Section 3 contains all the proofs followed by an Appendix.

\section{Main Results}   
In this section we shall be concerned with the multivariate setup  deriving  first an  extension of \netheo{th1} and then discussing  further some related conditional limit results. Given its importance in application we shall consider also  the approximation of MCTE. In the last subsection the three dimensional case will be briefly explored.\\
 In our notation below bold lower case symbols are column vectors in $\R^d$. 
The Hadamard product $r \x$ stands for the vector $(r x_1 \ldot r x_d)$ where $r\inr, \x=(x_1 \ldot x_d)^\top \inr^d$.  All other  operations with vectors are defined as usual, component-wise. For instance $\vk a \vk x$ is the vector $(a_1 x_1 \ldot a_d x_d)^\top $ for any $\vk a, \vk x\inr^d$ and $\vk x \ge \vk a$ means that $x_i \ge a_i, i\le d$.   

\subsection{Approximation of MME and MES}  
Let in the following  $\vk Z=(Z_1 \ldot Z_d)$ be a $d$-dimensional Gaussian random vector with mean $\vk \mu$. As in the bivariate case we  define MME for give level $p\in (0,1)$   by 
$$ E(p)= \E{(Z_1- A_p)_+ \lvert  Z_2> VaR_{Z_2}(p) \ldot  Z_d > VaR_{Z_d}(p)   },$$
with $A_p= \sum_{i=1}^{d-1} a_i VaR_{Z_{i+1}}(p)$ where $a_i$'s are given   constants.  
Writing  $\sigma_i^2$ for the variance of $Z_i$ we have thus   
\bqny{ 
E(p)&=& \sigma_1 \E{(X_1 - ( A_p   -\mu_1)/\sigma_1 )_+ \lvert  X_2> VaR_{X_2}(p) \ldot  X_d > VaR_{X_d}(p)   }\\
&=& \sigma_1 \E{(X_1 - (\sum_{i=1}^{d-1} a_i ( \sigma_{i+1} u_p + \mu_{i+1})   -\mu_1)/\sigma_1 )_+ \lvert  X_2> u_p  \ldot X_d > u_p    },
}
with $\X=(X_1 \ldot X_d) $ a centered Gaussian random vector with covariance matrix $\Sigma$ equal to the correlation matrix of $\vk  Z$ and 
$u_p= \Phi^{-1}(p)$. For  notational simplicity, throughout this paper random vectors are row vectors and therefore we do not use  the transpose sign.\\
  Consequently, without loss of generality we shall determine next the asymptotics of  
$$ E(\vk c,u)= \E{(X_1 -  c_1u - \mu)_+ \lvert  X_2> c_2 u \ldot  X_d > c_d u   }$$
as $u\to \IF$ for given  $\vk c=(c_1 \ldot c_d)^\top,\mu$ assuming that $\Sigma$ is a non-singular correlation matrix.\\
 In the two-dimensional setup the aimed approximation  can be obtained without discussing a closely related and crucial quadratic optimisation problem. However, in the higher dimensional settings we need to solve the following quadratic programming problem 
$ \Pi_\Sigma(\vk c)$: determine the minimum of $\vk x^\top \Sigma^{-1}\x$ subject to $\x \ge \vk c$ for given $\vk c\inr^d\setminus (-\IF, 0]^d$ with   solution $\tilde{\vk c}$. The reason for discussing $\Pi_\Sigma(\vk c)$ is that our investigation is closely related to the  asymptotic tail behaviour as $u\to \IF$ of 
$\pk{\X> \vk c u}$. In view of  \cite{ENJH02} (see below   \nelem{thmL}) the aforementioned asymptotic tail behaviour  is solely determined by $\Pi_\Sigma(\vk c)$. 

  In view of  \nelem{prop1} in Appendix   we have that $\vk{\tilde c}$ exists,  is unique and  there exists a unique  index set $I \subset \{1 \ldot d \}$ with $m\ge 1$ elements  such that
 \bqn{\label{cc} \tilde{\vk c}_I= \vk c_I, \quad \tilde{\vk c}_{I^c}= \Sigma_{I^cI}(\Sigma_{II})^{-1} \vk c_I \ge  \vk c_{I^c}, \quad 
 	\tilde{\vk c}^\top  \Sigma^{-1} \tilde{\vk c} = \vk c_I^\top 
 	(\Sigma_{II})^{-1} \vk c_I>0,
}
where $I^c= \{1 \ldot d \} \setminus I$; note in passing that $I^c$ 
can be empty.\\
Throughout this paper   $\Sigma_{IJ}$ is the matrix obtained by $\Sigma$  keeping the rows and columns with indices in $I$ and $J$,  respectively and  similar notation applies for vectors. 

 Denote  next by $L \subset \{1 \ldot d \}$ the maximal index set that contains $I$ such that 
  $\tilde{\vk c}_L=\vk c_L$. We have  by \nelem{prop1}  that   
  $$\vk c^\top  \Sigma^{-1} \vk c = \vk c_L^\top 
 (\Sigma_{LL})^{-1} \vk c_L=  \vk c_I^\top 
 (\Sigma_{II})^{-1} \vk c_I$$
 and moreover 
 \bqny{  \label{raki} 
 h_i= \vk c_I^\top (\Sigma_{II})^{-1} \vk e_i > 0, \quad \forall i\in I	,
}	
where $\vk e_i$ is the unit vector    in $\R^m$  with all components equal to 0 apart from the $i$th component equal to 1. 
 Denote by $L^c$ the complement of index set $L$ with respect to $\{1 \ldot d\}$. \\
For illustration purposes, we discuss briefly  the case $d=2$. 
Consider therefore $\Sigma$ to be a correlation matrix with off diagonal elements equal $\rho\in (-1,1)$ and let $\vk c= (1,c)^\top$. If $c\in ( \rho,1) $, then  $\tilde{\vk c}= \vk c$ and hence $I=L=\{1,2\}$ implying that  $I^c,L^c$ are empty. The assumption $c=\rho$ yields  
$$I=\{1\}, \quad L=\{1 , 2\},$$
whereas supposing that $c< \rho$ implies  $\tilde{\vk c}= (1, \rho)^\top$ and $I=L=\{1\}$.   

Below we write $\vk z_{-1}$ instead of $\vk z_I$ with $I=\{2 \ldot d \}$ for any $\vk z\inr^d$.  We present  next the approximation of $E(\vk c,u)$.

\BT Let $\vk c,\mu$ be two given constants and let $\Sigma$ be the  non-singular covariance matrix of the centered Gaussian random vector $\X$.  Let $I,L$  be the index sets identified by $\Pi_\Sigma(\vk c),$ where $\vk c \inr^d$ has at least one positive component. \\
i) If $1\in I$, then we have 
\bqn{ E(\vk c,u) \sim  \frac 1 {\vk c_I^\top (\Sigma_{II})^{-1} \vk e_1} \frac{\pk{ X_1> c_1 u+ \mu, 	\vk X_{-1} > \vk c_{-1} u}}{u  \pk{ 
			\vk X_{-1} > \vk c_{-1} u } } \to 0, \quad u\to \IF. \label{27}
}
ii) If $1 \in L \setminus I$, then $c_1 =  (  \Sigma_{I^cI} (\Sigma_{II})^{-1} \vk c _I)_1$ and further  
\bqn{ \label{meinart}
	\limit{u} E(\vk c,u) &=&  \E{(Y- \mu)_+}\in (0,\IF),
	 \label{23}}
where $Y$ has survival function $\bar G(x)= \pk{X_1> x\lvert \X_I= \vk 0_I}$ if $L=I \cup \{1\}$ and if $N^*=L \setminus (I \cup \{1\})$ is non-empty 
\bqn{\label{barG}
	\bar  G(x)=\frac{ \pk{ X_1 > x, \vk X_{N^*}> \vk 0_{N^*}\lvert \X _I = \vk 0_I} }{\pk{ \vk X_{N^*}> \vk 0_{N^*}\lvert \X _I = \vk 0_I}}, \quad   x\inr.
}
iii) If $1\in L^c$, then as $u\to \IF$
\bqn{ E(\vk c,u) \sim u((  \Sigma_{I^cI} (\Sigma_{II})^{-1} \vk c _I)_1 - c_1) \to \IF.
}
\label{thm2}  
\ET

\BRM i) The tail asymptotics of Gaussian random vectors is well-known, see below  \nelem{thmL} for a minor refinement. Hence  the exact asymptotic behaviour of 
$E(\vk c,u)$ in \eqref{27} can be explicitly calculated by approximating both  $\pk{ X_1> c_1 u+ \mu, 	\vk X_{-1} > \vk c_{-1} u}$ and 
	$\pk{ 	\vk X_{-1} > \vk c_{-1} u}$ as $u\to \IF$. \\
ii) As we demonstrate in the Appendix,  $E(\vk c,u)$ in \eqref{27} equals $o(e^{- \varepsilon u^2})$ for some small  $\varepsilon>0$.\\
 \ERM
In order to discuss the approximation of MES in this $d$-dimensional setting  we define 
\bqny{ 
	S(\vk c,u):=\E{ X_1 \lvert 	\vk X_{-1} > \vk c_{-1} u} &=& c_1 u + \E{ X_1- c_1u  \lvert 	\vk X_{-1} > \vk c_{-1} u}   \\
	&= :& c_1 u +  A(\vk c, u) , \quad \vk c= (c_1 \ldot c_d)^\top ,
}
where $\vk c\inr^d\setminus (- \IF, 0]^d$. Since we are interested in  the approximation  of 
	$\E{ X_1 \lvert \vk X_{-1} > \vk c_{-1} u } $ as $u\to \IF$, the natural question here is if we can determine $c_1$ such that $A(\vk c, u)$ is bounded for all large $u$.\\ In view of \cite{MR2397662}[Thm 5.1], we know that for particular choices of $\vk c$ the following convergence in distribution 
	\bqn{ 
		(X_1 - c_1 u) \lvert (	\vk X_{-1} > \vk c_{-1} u) \todis Y, \quad u\to \IF
\label{lum}
}
	 holds with $Y$ being a Gaussian or some truncated Gaussian random variable. The aforementioned result suggests that $\limit{u} A(\vk c, u)=\E{Y}$ could be valid, which then  for the specific choice of $c_1$ implies  
	\bqn{ \label{zez} 
	S(\vk c,u)- c_1u  \to \E{Y} , \quad u\to \IF.	
} 
Our next result shows that indeed \eqref{zez} holds. 

\BT \label{thm3} 
Let $\vk b= \vk c_{-1} $ have at least one positive component and let $\mathcal I, \mathcal L  $ be the  index sets corresponding to $\Pi_B(\vk b )$ with unique solution $\tilde{\vk b}$, where $B$ is the covariance matrix of $\vk X _{-1} $.  Suppose for simplicity that $\mathcal I=\{k \ldot d-1 \}$. Then \eqref{zez} holds with 
$$ 
c_1= \Sigma_{1, I} (\Sigma_{II})^{-1} \vk c_I, \quad I= \{ k+1 \ldot d\}.
$$ 
Moreover, for the above choice of $c_1$ \eqref{lum} is satisfied with $Y$ having survival function 
$\bar G(x)= \pk{ X_1> x \lvert  \vk X_{I} = \vk 0}$    
if $\mathcal L= \mathcal I$. In case that $\mathcal N= \mathcal L \setminus \mathcal I$ is non-empty, then $\bar G$ is given from \eqref{barG} with  $N^*= \mathcal N+1$.
\ET

\BRM  \label{remkoka} In the two dimensional setup $\vk b$ has only one element and thus $\mathcal I= \mathcal L$. Hence the limiting random variable 
$Y$ has $N(0, 1- \rho^2)$ distribution and therefore $\E{Y}=0$  confirming \eqref{eqe1}. 
If $\mathcal I\not= \mathcal L$,   then in general $\E{Y}$ does not equal  0. 
\ERM 

\subsection{Approximation of MCTE}
Another  interesting risk measure is the multivariate conditional tail expectation (abbreviated here as MCTE), which for elliptically symmetric random vectors  can be calculated explicitly, see \cite{MR3543047,MR3754572}. For a given random vector $\X=(X_1 \ldot X_d)$ with integrable components and given $\vk c\inr^d$ it is  defined by  
$$ M(\vk c, u)=\E{ X_1  \lvert \X > \vk c u }$$ for $u>0$ and $\vk c$ with at least one positive component. \\
 Note in passing that for any $\vk c,u$  and taking  for simplicity $\mu=0$ we have (hereafter where $\mathbb{I}(\cdot)$ denotes the indicator function)
 \bqny{
 E(\vk c,u)&=& 
\frac{ \E{ (X_1 -  c_1u )_+ \mathbb{I}(\X_{-1}> \vk c_{-1} u)  }}{\pk{\X_{-1}> \vk c_{-1} u}}\\
&=& \frac{ \E{ (X_1 -  c_1u )\mathbb{I}(X_1> c_1u) \mathbb{I}(\X_{-1}> \vk c_{-1} u) }}{\pk{\X_{-1}> \vk c_{-1} u} }\\
&=& \frac{\pk{\vk X> \vk c u} } {\pk{ \vk X_{-1}> \vk c_{-1} u}} 
\frac{ \E{ (X_1 -  c_1u ) \mathbb{I}( \X > \vk c u ) }}{\pk{\X > \vk c u} }
	\\
&=& \frac{\pk{\vk X> \vk c u} } {\pk{ \vk X_{-1}> \vk c_{-1} u}} 
\E{ (X_1 -  c_1u ) \lvert  \X > \vk c u ) } \\
 	&=:& r(u ) [M(\vk c, u)  - c_1 u],
} 
 where we assumed that $ \pk{\X> \vk cu} >0$.  In view of  \nelem{thmL}, under the assumption $iii)$ in \netheo{thm2} it follows 
   that  $\limit{u} r(u)=1$.  Consequently,  \netheo{thm2} implies 
\bqn{\label{ms}
	   M(\vk c, u) \sim  \Sigma_{1,I} (\Sigma_{II})^{-1} \vk c_I u, \quad u\to  \IF.
}
Under the assumption $ii)$ in \netheo{thm2} since by \nelem{thmL} we have $\limit{u} r(u)= C\in (0,\IF)$, then again \netheo{thm2} yields that for some $C_1>0$ that can be calculated explicitly 
\bqn{ \limit{u} [M(\vk c,u) - c_1u] =C_1, \quad u\to \IF.
}
Finally, under the assumptions of \netheo{thm2}, $i)$ we have that 
\bqn{
	\limit{u} u[M(\vk c,u) - c_1u] = \frac{1}{\vk c_I^\top (\Sigma_{II})^{-1} \vk e_1}>0, \quad u\to \IF.
}
 An intuition for the above approximations  comes from the conditional limit theorem derived in \cite{MR2397662}[Thm 5.1].  For instance if $1 \in I$ being the  index set related to $\Pi_\Sigma(\vk c)$ for some general $\vk c$ with at least one positive component, we have the convergence in distribution  
$$ u(X_1 - c_1 u) \bigl  \lvert ( \X > \vk c u ) \todis  \mathcal{E}, \quad u\to \IF, $$
where  $\mathcal{E}$ is an exponential random variable with mean $1/\vk c_I^\top (\Sigma_{II})^{-1} \vk e_1$.

The following result is new and gives a minor refinement of  \eqref{ms}.   

\BT Under the assumptions of \netheo{thm2} iii) we have $ \tilde c_1= \Sigma_{1,I} (\Sigma_{II})^{-1} \vk c_I>c_1$ 
\bqn{ (X_1 - \tilde c_1 u) \lvert \X > \vk c u \todis Y, \quad u\to \IF,
 }
where  $Y$ has survival function  $\bar G$ given in \netheo{thm2} with $N^*=L \setminus I.$  Moreover as $u\to \IF$ 
\bqn{ \label{besserM} 
	M(\vk c,u)- \tilde c_1 u \to \E{Y}.}
\label{thm4}
\ET 

\BRM 
 If $L=I$, then    $\E{Y}=0$ since $Y$ with survival function  $\bar G$ defined above is a centered Gaussian random variable.\\
 \ERM 

\subsection{Trivariate Case} In order to apply our results we need to determine  the index sets $I$ and $L$ related to the quadratic programming problem $\Pi_\Sigma(\vk c)$. The index set $I$ has $m\le d$ elements and it is possible that $m=1$ for given $\vk c$ with at least one positive component. If $X_1$ is independent of $\vk X_{-1}$, then it follows easily that $m\ge 2$ and $1 \in I$, whereas for the case $d=2$ and $c_1=c_2$ we have  $m=2$ and $I=L$. In general, $m=d$ if and only if the so-called Savage condition (see \cite{Savage,Ruben}) 
$$\Sigma^{-1} \vk c> \vk 0=(0 \ldot 0)^\top \inr^d$$
holds, which can be easily checked for given $\vk c$ and $\Sigma$. If the Savage condition does not hold,  then $m<d$ but the exact value of $m$ cannot be known without the knowledge of $\Sigma$ and $\vk c$. In the following we discuss in details the trivariate case  $\vk c= (1,1,1)^\top $ and $\Sigma$ is a non-singular correlation matrix with entries $\sigma_{ij},i,j\le 3$.\\  
First note that the Savage condition  is equivalent with 
\bqn{\label{arge} 
	1+ 2 \sigma -\sigma_{12}- \sigma_{13}- \sigma_{23}> 0, \quad \sigma= \min(\sigma_{12}, \sigma_{13}, \sigma_{23}),
}
which is equivalent with $m=3$ as mentioned above. Consequently, assuming \eqref{arge}, by statement $i)$ in \netheo{thm2}  
\bqny{
E(\vk c, u) \sim  
\frac{\sqrt{1-  \sigma_{23}} } {(1+ \sigma_{23})^{3/2} } \frac{1}{ \sqrt{2 \pi det(\Sigma)} (\vk c^\top  \Sigma^{-1} \vk e_1)^2 \prod_{i=2}^3 \vk c^\top  \Sigma^{-1} \vk e_i } \frac{1}{u^2}  e^{ - \frac{u^2}{2} [ ( \vk c + \mu \vk e_1)^\top \Sigma^{-1} (\vk c + \mu \vk e_1)/2 - 1/(1+ \sigma_{23})]}  
}
as $u\to \IF$, where $\vk e_i$'s are unit vectors in $\R^ d$ with $1$ in the $i$th coordinate and all other coordinates equal 0.\\
Suppose next that  
\eqref{arge} does not hold, i.e.,   
\bqny{ \label{wild}
	 1+ 2\sigma - \sigma_{12}- \sigma_{13}- \sigma_{23}\le 0
	}
and  $m=2$ since $m=1$ is impossible in the two dimensional setup when the coordinates of $\vk c$ are equal and positive.
  If \eqref{wild}  is satisfied with equality, then $L=\{1,2 ,3\}$. Assuming that  $\sigma_{12}\le \min( \sigma_{13},\sigma_{23})$ implies $I=\{1,2\}$ and thus $1\in I$ and the asymptotics of $E(\vk c, u)$ follows again from   statement $i)$ in \netheo{thm2}. 
The case $\sigma= \sigma_{13}$ is similar and therefore we assume next that $\sigma=\sigma_{23}$, which implies that $I=\{2,3\}$ and thus $1\not \in I$ and $1\in L$, provided that   $1+ \sigma_{23}= \sigma_{12}+ \sigma_{13}$. For this case, by \eqref{meinart} 
\bqn{ \limit{u} E(\vk c,u)= \E{(X_1- \mu)_+ \lvert X_2=0, X_3=0}.
	}
  Finally, if $\sigma_{12}+ \sigma_{13}- \sigma_{23}-1>0$, then $I=L$ and $1 \in L^c$. Hence by statement $iii)$ in \netheo{thm2}  
 \bqny{  E(\vk c,u) \sim   \frac{\sigma_{12} + \sigma_{13}- \sigma_{23}-1}{1+ \sigma_{23}}  u
 }  
and from  \eqref{besserM} 
$$ M(\vk c,u) -  \frac{\sigma_{12} + \sigma_{13}}{1+ \sigma_{23}}  u \to 0$$
as $u\to \IF$.
 
\section{Proofs} 

\prooftheo{th1} Let $(X_1,X_2)$ be jointly Gaussian with mean vector zero, correlation $\rho\in (-1,1)$ and set 
$$u:=u_p= VaR_{X_2}(p), \quad \beta=  \frac{\mu_2- \mu_1}{\sigma_1}, \quad c= \frac{\sigma_2}{\sigma_1}.$$
For any $u>0$ we have 
\bqny{ E(p) &=& \E{ (\sigma_1 X_1 + \mu_1 -  \sigma_2 u - \mu_2 )_+ \lvert  X_2 >  u     }\\
	&=& 
	\sigma_1\E*{ \Bigl( X_1  -    \frac{ \sigma_2}{\sigma_1} u    - \frac{\mu_2- \mu_1}{\sigma_1}\Bigr )_+ \Bigl \lvert  X_2 >  u    }\\
	&=& \frac{ \sigma_1}{\pk{X_1> u}}\E*{ ( X_1  -    c u    - \beta  ) \mathbb{I}( X_1 > cu+ \beta,  X_2 >  u     } \\
	&=:&\frac{ \sigma_1}{\pk{X_1> u}}\theta_u \in (0,\IF).
} 
Let below $\varphi$ denote the pdf  
of $(X_1,X_2)$.\\
$i)$ First note that in this case $c \in ( \rho,1]$. Let $h_1^*,h_2^*$ be defined by 
\bqn{\label{frank}
	 h_1^*= \frac{c - \rho}{1- \rho^2}>0, \quad h_2^*= \frac{1- c \rho}{1- \rho^2}>0.
	}
Using the transformation
$$ s= c u+\beta + x/ u,\quad   t= u + y/u$$
for any $u>0$, we have further
\bqny{
	\lefteqn{\theta_u =\int_{cu+ \beta}^\IF \int_u^\IF (s- cu- \beta) \varphi(s,t) ds dt}\\
	&=& 
	{u^{-3}}\int_{0}^\IF \int_0^\IF x \varphi(cu+ \beta + x/u,u+ y/u) dx dy\\
	&=:&  	{u^{-3}}\varphi(cu+ \beta , u)
	\int_{0}^\IF \int_0^\IF x \expon{- h_1^* x- h_2^* y }  \psi_u(x,y) dx dy.
}
After some calculations  for any $x,y$ positive  we obtain
\bqn{ \label{psi} \limit{u}  \psi_u(x,y)=1}
and further for all $\varepsilon >0$ sufficiently small and all $u$ large $\psi_u(x,y) \le e^{ \varepsilon (x+ y)}$. Consequently, since $h_1^*,h_2^*$ are positive,  applying the dominated convergence theorem 
we obtain  
\bqny{ \theta_u &\sim &  {u^{-3}}   \varphi(cu+ \beta , u) 
	\int_{0}^\IF \int_0^\IF x \expon{-h_1^*x - h_2^* y }  dx dy \\
	&=&  \frac{1}{ (h_1^*)^2 h_2^*} {u^{-3}} \varphi(cu+ \beta, u), \quad u\to \IF,
}
hence the claim follows.  \\
$ii)$ If $c= \rho$ the above transformation cannot be used since then $h_1^*= 0$ and the limiting integral is not finite. We use another transformation, namely
$$ s= \rho u+\beta + x ,  \quad t= u + y/u$$
for any $u>0$. Consequently, we have  
\bqny{
	\lefteqn{\theta_u = u^{-1} \int_{0}^\IF \int_0^\IF x \varphi (\rho u+ \beta + x,u+ y/u) dx dy} \\
	&=:& u^{-1}   \varphi(\rho u, u) 
	\int_{0}^\IF \int_0^\IF x e^{ - \frac{(x+ \beta)^2}{2 (1- \rho^2)} - y  } \psi_u(x,y) dx dy.
}
By the definition of $\varphi$  
\bqn{\label{hgoja}
	u^{-1}   \varphi(\rho u, u)  \sim \frac{1}{2 \pi \sqrt{1- \rho^2}} u^{-1} e^{-u^2/2},  \quad 
	\pk{X_1> u} \sim u^{-1} e^{-u^2/2}/\sqrt{2 \pi}, \quad u\to \IF,
}
where the second approximation is a direct consequence of the well-known Mill's ratio asymptotics. Clearly \eqref{psi} holds and the domination of the integrand follows easily. Hence by  the dominated convergence theorem   as $u\to \IF$
\bqny{
	\theta_u & \sim & u^{-1} \varphi(\rho u, u)
	\int_{0}^\IF \int_0^\IF x e^{ - \frac{(x+\beta)^2}{2 (1- \rho^2)} - y  }  dx dy \\
	&\sim &  \frac{1}{\sqrt{2 \pi (1- \rho^2)}} 
	\int_{\beta}^\IF  (x-\beta) e^{ - \frac{x^2}{2 (1- \rho^2)}   }  dx  \pk{X_1> u} \\
	&	= &\E{( \sqrt{1- \rho^2}X_1 - \beta)_+}\pk{X_1> u}.
}
Since for any $a>0, b\inr$ 
\bqn{\label{ab} 
	\E{( a X_1 - b)_+} = a\Phi'( b/a) -
b[1- \Phi( b/a)]
}
%
the claim follows. \\
$iii)$ If $c< \rho$, then     
\bqny{ 
	\pk{X_1> cu + \beta , X_2> u} &=& u^{-1} \int_{0}^\IF \pk{  \sqrt{1- \rho^2} X_1> (c- \rho)u - \rho x/u+\beta } \frac{1}{\sqrt{2 \pi}} e^{- (u+ x/u )^2/2} \, dx\\
	&\sim & \pk{X_1> u}, \quad u \to \IF.
} 
Next,  using the same transformation as for the case $c=\rho$  gives letting  $u\to \IF$ 
\bqny{
	\theta_u &=&  \int_{cu+\beta}^\IF \int_u^\IF (x+ (\rho -c)u 	 - \rho u - \beta ) \varphi(x,y) dx dy \\ 
	&=& 	(\rho -c)u \pk{X_1> cu+ \beta  , X_2> u}+	 \int_{cu+ \beta}^\IF \int_u^\IF (x- \rho u - \beta ) \varphi(x,y) dx dy \\
	&\sim&  (\rho -c) u\pk{X_1> u}  + 	
	u^{-1} \int_{(c - \rho)u }^\IF \int_0^\IF x \varphi(\rho u+ \beta +x,u+ y/u) dx dy  \\
	&= & 
	(\rho -c) u\pk{X_1> u}  + 	
	u^{-1} \varphi(\rho u, u)  \int_{(c - \rho)u }^\IF \int_0^\IF x e^{ - \frac{(x+\beta)^2}{2 (1- \rho^2)} - y  } \psi_u(x,y) dx dy.
}
As above, by \eqref{psi} and $ \limit{u} (c-\rho)u= - \IF$ 
$$ \limit{u} \int_{(c - \rho)u }^\IF \int_0^\IF x e^{ - \frac{(x+ \beta)^2}{2 (1- \rho^2)} - y  } \psi_u(x,y) dx dy = 
\int_{\R } \int_0^\IF x e^{ - \frac{(x+ \beta)^2}{2 (1- \rho^2)} - y  }  dx dy=0.
$$
Utilising further \eqref{hgoja}  we obtain   
$$ \theta_u \sim (\rho-c) u\pk{X_1> u}, \quad u\to \IF$$
establishing the claim.\\
$iv)$ Since  $c\ge 1/\rho$, then  $h_2^*$ defined in \eqref{frank} is non-positive. Hence we need to use another transform, namely 
$$ s=  c u  +\beta + x/u ,  \quad t=  c \rho u + y$$
for any $u>0$. Consequently, for any $u>0$    
\bqny{
	\theta_u &=& u^{-2} \int_{0}^\IF \int_{ (1- c \rho) u} ^\IF x \varphi (c u+ \beta + x/u, c \rho u+ y) dx dy \\
	&=:& {(cu)}^{-2}   \varphi(c u, c \rho u) e^{-\beta cu} 
	\int_{0}^\IF\int_{ (1- c \rho) u}^\IF x e^{-x} e^{ - \frac{ y^2 - 2 \rho \beta y + \beta^2}{2 (1- \rho^2)}}  \psi_u(x,y) dx dy,
}
where $\psi_u(x,y) \to 1$ as $u\to \IF$. The domination of the integrad follows easily, hence applying the dominated convergence theorem  and \eqref{hgoja},  for $c=1/\rho$ 
\bqny{
	\theta_u & \sim &  {(cu)}^{-2} \varphi(c u, c \rho u) e^{  -\frac{ \beta^2}{2 } -\beta cu} 	\int_0^\IF  \int_{0}^\IF x e^{-x}  e^{ - \frac{(y- \rho \beta )^2}{2 (1- \rho^2)}   }  dx dy \\
	&=&  {(cu)}^{-2} \varphi(c u, c \rho u) e^{  -\frac{ \beta^2}{2 } -\beta cu}  \sqrt{ 2 \pi (1- \rho^2)} [1- \Phi( -  \rho \beta/\sqrt{1- \rho^2}) ]\\
	&\sim & {(cu)}^{-1} e^{  -\frac{ \beta^2}{2 } -\beta cu}   {\Phi( \rho \beta/\sqrt{1- \rho^2})} [1- \Phi(cu)]  
}
as $u\to \IF$. If $c> 1/\rho$, then 
\bqny{
	\theta_u & \sim &   {(cu)}^{-2} 
	\varphi(c u, c \rho u) e^{  -\frac{ \beta^2}{2 } -\beta cu}
	\int_{\R }  e^{ - \frac{(y- \rho \beta )^2}{2 (1- \rho^2)}   }   dy \\
	&\sim &  {(cu)}^{-1} e^{  -\frac{ \beta^2}{2 } -\beta cu}[1- \Phi(cu)],  \quad u\to \IF,
}
hence the claim follows.\\
$v)$ First note that for any $p \in (0,1)$ and $u:=u_p= VaR_{Z_2}(p)$ 
$$ S(p)=\mu_1+ \E{( Z_1- \mu_1)  \lvert Z_2> VaR_{Z_2}(p) } = \mu_1+ \sigma_1 \rho u + \sigma_1  \E{ X_1- \rho u \lvert X_2> u}.$$ 
As above we have 
\bqny{ 
	\E{ X_1- \rho u \lvert X_2> u} &=& \frac{1}{\pk{X_1> u}} \int_{x \inr, y> u} (x- \rho u) \varphi(x,y) dx dy\\
	&=&   \frac{\varphi(\rho u, u)}{u\pk{X_1> u}} \int_{x \inr, y> 0} x \varphi(\rho  u+ x, u+y/u)/\varphi(\rho u, u) dx dy\\
	&\sim &  \frac{1}{\sqrt{2 \pi (1- \rho^2)}} \int_{x \inr, y> 0} x \varphi(\rho  u+ x, u+y/u)/\varphi(\rho u, u) dx dy\\
	&\sim &  \frac{1}{\sqrt{2 \pi (1- \rho^2)}} \int_{x \inr, y> 0} x e^{-\frac{x^2}{2 (1- \rho^2)} - y}  dx dy\\
	&=&0
}
as $u\to \IF$,  establishing thus  the claim. 
\QED

\prooftheo{thm2} 
The proof is driven by the tail asymptotics of Gaussian random vectors derived in \cite{ENJH02}.  
As therein the index set $I$ is also crucial for the derivation of the asymptotics of $E(\vk c, u)$, since  the tail asymptotics of $\pk{\X > \vk c u}$ is up to a pre-factor the same as that of $\pk{\X_I> \vk c_I u}$ as $u\to \IF$.  The components with indices in the set $L \setminus I$ influence the asymptotics only by the pre-factor, whereas the components with indices in the set $K:=L^c$ are not important. For these reasons we have three different cases which shall be dealt with separately.\\

Set next for any $u>0$ 
$$E^*(u)=\pk{\vk X_{-1} > \vk c_{-1} u }E(\vk c,u)$$
and write $\varphi$ for the pdf of $\vk X$.\\
$i)$ When $1 \in I$, then  $\tilde{c}_1=c_1$. Hence for any $u$ positive  
\bqny{ E^*(u)
&=& \int_{ \vk s > \vk c u+ \mu \vk e^*_1}( s_1- c_1u- \mu)_+ \varphi(\vk s) d \vk s\\
	&=& \frac{1}{u^{m+1}} \int_{ \vk x> \bar{\vk u} (\vk c u- \tilde{\vk c}u)  }  x_1  \varphi(\tilde {\vk c} u+ \vk x/\bar {\vk u}  + \mu  \vk e^*_1     ) d \vk x,
}	
where $\bar {\vk u}$ has all components with indices  in $I$ equal to $u$ and otherwise equal to 1 and $\vk e^*_1$ has all components equal to 0 apart from the first component equal to 1.  Recall that $m$ stands for the number of the elements of the index set $I$ which cannot be empty. Using further 
\eqref{eq:new} (set next $J=I^c=\{ 1 \ldot d \} \setminus I$ and assume for simplicity that $J$ is not empty) we have  

\bqn{ \label{domA} 
	\lefteqn{	(\tilde{\vk c}u + \vk x/\bar{\vk u}  + \mu  \vk e^*_1     ) ^\top \Sigma^{-1} (\tilde{\vk c}u + \vk x/\bar{\vk u}  + \mu  \vk e^*_1     )  }\notag \\
	&=&  
	(\tilde{\vk c}u   + \mu  \vk e^*_1     ) ^\top \Sigma^{-1} (\tilde{\vk c} u + \mu  \vk e^*_1     )  
	+ 2 	u\tilde{\vk c}^\top \Sigma^{-1} \vk x/ \bar{\vk u}  + 2\mu ( {\vk e}^*_1)    ^\top \Sigma^{-1}\vk x/\bar 	{\vk u}+ (\vk x/\bar{\vk u})^\top  \Sigma^{-1} \vk x/\bar{\vk u}.
}
By the properties of $\tilde {\vk c}$ (see equation \eqref{eq:new} in \nelem{prop1}) for any $u\not=0,\x\inr^d$ 
$$    	u\tilde{\vk c}^\top \Sigma^{-1} \vk x/ \bar{\vk u} = 	\tilde{\vk c}_I^\top (\Sigma_{II})^{-1} \vk x_I .$$
Hence since $1 \in I$ implies $ (\vk e^*_1) ^\top \Sigma^{-1}\vk x/\bar{\vk u} = O(1/u)$ as $u\to \IF$, then by \eqref{domA} 
$$ 
\varphi(\tilde {\vk c} u+ \vk x/\tilde{\vk u}  + \mu  \vk e^*_1     )   =  \varphi(\tilde {\vk c} u  + \mu  \vk e^*_1  )\psi_u(\vk x) e^{ - 	{\vk c}_I  (\Sigma_{II})^{-1}\vk x_I - \vk x_J^\top  (\Sigma^{-1})_{JJ } \vk x_J/2 },
$$
where   $\limit{u}\psi_u(\vk x) = 1$ for any $\x \in \R^d$.  Using the fact that  $\Sigma^{-1}$ is positive definite and ${\vk c}_I^\top   (\Sigma_{II})^{-1} > \vk 0_I$  for any $\x\inr^d$ with $\x_I> \vk 0_I$  we obtain that
\bqn{\label{domC} 2 	{\vk c}_I^\top   (\Sigma_{II})^{-1} \vk x_I+  2\mu ( {\vk e}^*_1)    ^\top \Sigma^{-1}\vk x/\bar 	{\vk u}+ (\vk x/\bar{\vk u})^\top  \Sigma^{-1} \vk x/\bar{\vk u} \le C ( \vk 1 _I^\top \vk x_I + \x^\top_J \x_J)
}
holds for all large $u$ and some positive constant  $C$. Using thus  the dominated convergence theorem (recall  $\tilde c_i > c_i $ for any $i\in K=L^c$) we obtain 
\bqny{ \lefteqn{ E^*(u)}\\
	&=& \frac{1}{u^{m+1}} \varphi (\tilde {\vk c} u  + \mu  \vk e^*_1  )  \int_{ \vk x_L > \vk 0_L, x_i >u (c_i- \tilde c_i), i \in K }  x_1   \psi_u(\vk x) 
	e^{ - 	{\vk c}_I  (\Sigma_{II})^{-1}\vk x_I - \vk x_J^\top  (\Sigma^{-1})_{JJ } \vk x_J/2 } d \vk x \\
	&\sim & \frac{1}{u^{m+1}} \varphi(\tilde {\vk c} u  + \mu  \vk e^*_1  )  \int_{ \vk x_L > \vk 0_L, x_i \in \R, i\in K} 
	 x_1   e^{ - 	{\vk c}_I  (\Sigma_{II})^{-1}\vk x_I - \vk x_J^\top  (\Sigma^{-1})_{JJ } \vk x_J/2 } d \vk x\\
	&= & \frac{1}{h_1 u} \frac{1}{u^{m}} \varphi(\tilde {\vk c} u  + \mu  \vk e^*_1  )  \frac{1}{\prod_{i\in I}h_i}  
	 \int_{ x_{i}> 0, i\in L\setminus I , x_i \in \R, i\in K} 
	  e^{ - 	 \vk x_J^\top  (\Sigma^{-1})_{JJ } \vk x_J/2 } d \vk x_J,
}
where $h_i= \vk c_I^\top (\Sigma_{II})^{-1} \vk e_i>0$ with  $\vk e_i$ the $i$th unit vector in $\R^{m}$ with $m$ the number of elements of the index set $I$. 
Since $1 \in I$, applying \eqref{sh} in \nelem{thmL} yields  
$$ E^*(u) \sim  (u h_1)^{-1} \pk{ \X>   \vk c u + \mu \vk e_1^*}, \quad u\to \IF , $$
hence the claim follows by the definition of $E^*(u)$.\\   
$ii)$ In view of \nelem{thmL}, the asymptotics of $\pk{ X_1 > c_1 u + x, \vk X_{-1}> \vk c_{-1} u }$ and that of 
$\pk{\vk X_{-1}> \vk c_{-1} u }$ as $u\to \IF$ are up to the pre-factor the same. It follows easily that 
$ Y_u:= (X_1 - c_1 u) \lvert \X_{-1}> \vk c_{-1}u $ converges in distribution as $u\to \IF$ 
to a random variable $Y$ which has survival function $\pk{X_1> x \lvert \X_I = \vk 0_I}$ if $L\setminus I= \{1\}$ and 
when $N^*=L \setminus (I \cup \{1\})$ is non-empty, then $Y$ has survival function 
$$ \frac{\pk{X_1> x, \X_{N^*}> \vk 0_{N^*} \lvert \X_I = \vk 0_I} }{\pk{\X_{N^*}> \vk 0_{N^*} \lvert \X_I = \vk 0_I}}, \quad x\inr.$$
In case that $(Y_u- \mu)_+, u>0$ is uniformly integrable, then 
$$ \limit{u} E(\vk u,c) = \E{(Y- \mu)_+}.$$
We show next the above convergence directly, which in turn implies the uniform integrability mentioned above. Since $1\in L \setminus I$ we still have that $\tilde c_1=c_1$ and as above 
\bqny{ E^*(u)&=& \int_{ \vk s > \vk c u+ \mu \vk e^*_1 }( s_1- c_1u- \mu)_+ \varphi(\vk s) d \vk s\\
	&=& \frac{1}{u^{m}} \int_{ \x >  \bar{\vk u} ( \vk c u -  \tilde{\vk c}u ) }  x_1  \varphi(\tilde {\vk c} u+ \vk x/\bar {\vk u}  + \mu  \vk e^*_1     ) d \vk x.
}	
Next, since $1 \not \in I$ i.e., $1 \in J:={I^c}$ by \eqref{domA} 
\bqny{ 
	\lefteqn{	(\tilde{\vk c}u + \vk x/\bar{\vk u}  + \mu  \vk e^*_1     ) ^\top \Sigma^{-1} (\tilde{\vk c}u + \vk x/\bar{\vk u}  + \mu  \vk e^*_1     )  }\\
	&=& (\tilde{\vk c} u   + \mu  \vk e^*_1     ) ^\top \Sigma^{-1} (\tilde{\vk c} u  + \mu  \vk e^*_1     )  
	+ 2 	{\vk c}_I^\top   (\Sigma_{II})^{-1} \vk x_I  + 2 \mu (\Sigma^{-1})_{1,J} \x_{J}+  \vk x_{J}^\top  (\Sigma^{-1})_{JJ } \vk x_{J} + O(u^{-1})
}
as $u\to \IF$.   
    Consequently, in view of \eqref{domC}, we can apply the dominated convergence theorem to obtain (set $N= L \setminus I$, write $k$ for the number of elements of the index set $K=L^c=\{1 \ldot d \} \setminus L$ and recall that $\tilde c_i> c_i, i\in K$)
\bqny{ E^*( u)
	&\sim & \frac{1}{u^{m}} \varphi(\tilde {\vk c} u +    \mu  \vk e^*_1 )  \int_{ \vk x_L > \vk 0_L, x_i >u (c_i- \tilde c_i), i \in K }  x_1  	e^{ - 	{\vk c}_I  (\Sigma_{II})^{-1}\vk x_I - \vk x_J^\top  (\Sigma^{-1})_{JJ } \vk x_J/2 - \mu (\Sigma^{-1})_{1,J} \x_J} d \vk x \\
	&= & \frac{1}{u^{m}} \varphi(\tilde {\vk c} u+\mu  \vk e^*_1    )  \frac{1}{\prod_{i\in I}h_i}
	 \int_{ \x_{N}> \vk 0_N, \vk x_K \in \R^{k}}   
	x_1 e^{ -  \vk x_J^\top  (\Sigma^{-1})_{JJ } \vk x_J/2 - \mu (\Sigma^{-1})_{1,J} \x_J} d \vk x_J
}
as $u\to \IF$.  With similar calculations 
\bqny{ \pk{ \X > \vk c u +   \mu \vk e^*_1}
	&\sim & \frac{1}{u^{m}} \varphi(\tilde {\vk c} u+\mu  \vk e^*_1    )  \frac{1}{\prod_{i\in I}h_i}
	\int_{ \x_{N}> \vk 0_N, \vk x_K \in \R^{k}}   
	 e^{ -  \vk x_J^\top  (\Sigma^{-1})_{JJ } \vk x_J/2 - \mu (\Sigma^{-1})_{1,J} \x_J} d \vk x_J
}	 
as $u\to \IF$. Since $1 \in J$, by  \nelem{thmL} 
$$ \limit{u} \frac{ \pk{ \X > \vk c u +   \mu \vk e^*_1}}{ \pk{\vk X_{-1} > \vk c_{-1} u }} = C_1$$
for some $C_1>0$  which can be calculated explicitly, hence the claim follows. \\
$iii)$ When $1 \in L^c$, then $\tilde c_1> c_1$ implying  
\bqny{ E^*(u)&=& \int_{\vk  s >  \vk c u+ \mu \vk e^*_1}( s_1- \tilde c_1 u  + (\tilde c_1 - c_1)u- \mu) \varphi(\vk s) d \vk s\\
	&=&(\tilde c_1 - c_1)u \pk{ \vk X > \vk c u + \mu \vk e^*_1     }  + \int_{ \vk s > \vk c u + \mu \vk e^*_1 }( s_1- \tilde c_1 u  - \mu) \varphi(\vk s) d \vk s.
}	
It follows easily that 
\bqny{ E^*(u)&\sim & (\tilde c_1 - c_1)u \pk{ \vk X > \vk c u+ \mu \vk e^*_1   }, \quad u \to \IF
}	
and further by \nelem{thmL}  
$$ \pk{ \vk X > \vk c u+ \mu \vk e^*_1    } \sim \pk{ \vk X_{-1}> \vk c_{-1} u    }, \quad u \to \IF,$$
hence the proof is complete. 
\QED

\prooftheo{thm3} We show first the conditional convergence in \eqref{lum}. Let $\tilde{\vk b}$ be the solution of the quadratic programming problem $\Pi_{B}(\vk b)$ with corresponding index set $\mathcal I=\{k  \ldot d-1 \}$ and let $\mathcal L$ be the set of indices such that $\tilde{b}_i=b_i$. Recall that $\vk b= \vk c_{-1}$ is a $(d-1)$-dimensional vector. 
Let  $I=\{k+1 \ldot d\}$ and set $c_1= \Sigma_{1, I} (\Sigma_{II})^{-1} \vk c_I$. By the definition of $\mathcal I$ and $I$ we have that 
$(\Sigma_{II})^{-1} \vk c_{I
}> \vk 0_{ I}$ and $ \tilde{\vk c}_{J^*}=\Sigma_{J^*I}(\Sigma_{II})^{-1}\vk c_I$ where $J^*=\{2 \ldot k\}$ being empty if $k=1$. Note that we agree that when index sets are empty, the defined relationships should be ignored.   Let $\tilde{\vk c}$ be such that $\tilde c_1=c_1= \Sigma_{1, I} (\Sigma_{II})^{-1} \vk c_I$ and $\tilde{\vk c}_{-1}=\tilde{\vk b}$.  Setting $J= \{1\} \cup J^*$ we have that 
$\tilde{\vk c}_{J}=\Sigma_{JI}(\Sigma_{II})^{-1}\vk c_I$. Consequently, since $(\Sigma_{II})^{-1} \vk c_{I
}> \vk 0_{ I}$ and $I\cup J=\{1 \ldot k\}$ by the converse statement in \nelem{prop1} we have that $\tilde{\vk c}$ 
is the unique solution of $\Pi_{\Sigma} (\vk c) $. From the aforementioned proposition $I$ is  the index set that determines  the unique solution $\tilde{\vk c}$. \\
 In order to show \eqref{lum} we need to determine the asymptotics as $u\to \IF$ of 
$$ \pk{\vk X > \vk c u+ x \vk e^*_1 }/\pk{\vk X_{-1} >\vk  c_{-1} u }$$ for any $x\inr$.  If $\mathcal L = \mathcal I$, then  
$L= \{1 \} \cup I$ (since $\tilde c_1=c_1$) and thus by  \nelem{thmL} we have that \eqref{lum} holds with $Y$ having the same distribution as $X_1 \lvert \vk X_I = \vk 0_I $. 
The case that $\mathcal N=\mathcal L \setminus \mathcal I$ is not empty   follows from \cite{MR2397662}[Corr 5.2]. Indeed, the tail asymptotics of the denominator and the nominator are the same up to some positive constant since the $I$ index sets of the corresponding quadratic programming problems are the same. The ratio of those constants is (set $N^*= \mathcal N+1$)
$$ \frac{ \pk{  X_1 > x,  \vk X_{ N^*} > \vk 0_{ N^*} \lvert \vk X_{I}  = \vk 0_I  }} 
{\pk{   \vk X_{ N^*}> \vk 0_{ N^*} \lvert \vk X_{I}= \vk 0 _I  }   }
 $$
and thus \eqref{lum} holds. The proof of  \eqref{zez} follows by  calculating  the asymptotics of 
$$ \E{ (X_1 - c_1 u) \mathbb{I}( \vk X_{-1}> \vk c_{-1}u )},
$$ 
which is established similarly to the proof of statement $ii)$ in  \netheo{thm2} and therefore we omit the details. 
\QED 

\prooftheo{thm4} Let $I,L$ denote the unique index sets defined from the solution of the quadratic programming problem $\Pi_\Sigma(\vk c)$. Suppose first that $N= L\setminus I$ is not empty. By the assumptions $1 \not \in N \cup I$. Let $\tilde{\vk {a}}$ be the unique solution of 
 $\Pi_{\Sigma}( \vk a), \vk a= (\tilde c_1, c_2 \ldot c_d)^\top$. The corresponding index set $I$ (write this as $I_{\vk a})$ includes $I$ since $1\not \in I$.  But we cannot have $1 \in I_{\vk a}$, i.e., $(\Sigma_{I_{\vk a}I_{\vk a}})^{-1} \vk a_{I_{\vk a}}> \vk 0_{I_{\vk a}}$ since this contradicts the definition of $a_1= \tilde c_1> c_1$. Consequently, $1$ belongs to the index set $L_{\vk a}$ of all indices $i\le d$ such that $\tilde a_i= a_i$. Next, for any $x\inr$ using \nelem{thmL} and \nelem{prop1}  we have 
\bqny{ \limit{u} \frac{ \pk{ X_1> \tilde c_1 u+ x, \vk X_{-1} >  \vk c_{-1}  u  }}{ \pk{\X > \vk c u}} =
	\frac{ \pk{ X_1> x, \X_N > \vk 0_N \lvert \vk X_I= \vk 0_I} }{\pk{  \X_N > \vk 0_N \lvert \vk X_I= \vk 0_I}} =: \bar G(x), \quad x\inr,} 
where for the asymptotics of the denominator we used the fact that $1\in L^c$, i.e.,  $\tilde c_1> c_1$. 
 If $I=L$, then $\bar G(x)= \pk{X_1> x \lvert \vk X_I= \vk 0_I}$. Consequently, $Y$ has the claimed survival function  $\bar G$.
  The second claim follows easily and therefore we omit the proof. 	
\QED

\section{Appendix}

\def\b{\vk b}
\def\tilb{\tilde{\vk b}}

\BEL \label{prop1} Let $\Sigma$ be a $d\times d$ positive definite matrix and let $\b \inr^d \setminus (-\IF, 0]^d $.
The quadratic programming problem
$\Pi_\Sigma(\b)$: minimise $\x^\top \Sigma^{-1} \x$ under $\vk x \ge \b$ has a unique solution $\tilb$ and there exists a unique non-empty
index set $I\subseteq \{1\ldot d\}$ with $m\le d$ elements such that
\BQN \label{eq:IJ}
\tilb_{I}=\b_{I} ,  \quad (\Sigma_{II})^{-1} \b_{I}>\vk{0}_I
\EQN
and if $I^c :=\{ 1 \ldot d\}\setminus I \not=\emptyset$, then 
\bqn{ \label{eq:IJ2} 
	\tilb_{I^c} &=& \Sigma_{{I^c}I} (\Sigma_{II})^{-1} \b_{I}\ge \b_{I^c},
}
\bqn{
\label{eq:IJ3}
\min_{\x \ge  \b}\x^\top \Sigma^{-1} \x= \tilb ^\top \Sigma^{-1} \tilb
&=& \b_{I}^\top (\Sigma_{II})^{-1}\b_{I}>0
}
\BQN \label{eq:new}
\x^\top \Sigma^{-1} \tilb= \x_F^\top (\Sigma_{FF})^{-1} \b_F, \quad  \forall \x\inr^d
\EQN
for any index set $F$ of $\{1\ldot d \}$ containing $I$ and if $\b= (b \ldot b)^\top , b\in (0,\IF)$,
then $ 2 \le \abs{I} \le d$. Conversely, if for some non-empty index set  
$I \subset \{ 1 \ldot d \}$ we have 
$$(\Sigma_{II})^{-1}\b_I> \vk 0_I,  \quad  \Sigma_{{I^c}I} (\Sigma_{II})^{-1} \b_I \ge \vk b_{I^c},$$
then $\tilb$ with $\tilb_{I^c}= \Sigma_{{I^c}I} (\Sigma_{II})^{-1} \b_I, \tilb_I= \vk b_I$ is the solution of $\Pi_{\Sigma}(\vk b)$.
 
\EEL 

\prooflem{prop1} The claims in \eqref{eq:IJ}-\eqref{eq:IJ3} are  formulated  in  \cite{Rolski17}.  
Since by \eqref{eq:IJ2} we have $(\Sigma^{-1} \tilb)_M= \vk{0}_M$ for any $M\subset {I^c}$ (assuming ${I^c}$ is not empty) exactly as in proof of  
\cite{BischoffCMP}[Lem 4.1] we have for any $\vk x\inr^d$ and $F= \{1 \ldot d \} \setminus M$ 
\bqn{ (\vk x+ \tilb)^\top \Sigma^{-1} (\vk x+ \tilb) = 
	\vk x^\top \Sigma^{-1} \vk x + 2\vk x_{F}^\top (\Sigma_{FF })^{-1}  \tilb_{F} +\tilb_{F}^\top (\Sigma_{FF } )^{-1}  \tilb_{F} , \label{coni}
}
which implies that $\vk x^\top \Sigma^{-1}  \tilb =\vk x_{F}^\top (\Sigma_{FF })^{-1}  \tilb_{F}$ and thus \eqref{eq:new} holds. 

If for some non-empty index set $I$ we have $(\Sigma_{II})^{-1} \vk b_I> \vk 0_I$,  then  $\b_I= argmin_{\x_I \ge \b_I} \x_I^\top (\Sigma_{II})^{-1} \x_I$.  Since for any two non-overlapping index set $A,B, A\cup B= \{ 1 \ldot d\}$  (using Schur compliments)
\bqny{
\x^\top  \Sigma^{-1} \x =\x_A^\top (\Sigma_{AA})^{-1} \x_A  + 
(\x_B - \Sigma_{BA}(\Sigma_{AA})^{-1}  \x_A) ^\top (\Sigma^{-1})_{BB}  (\x_B - \Sigma_{BA}(\Sigma_{AA})^{-1}  \x_A),\quad \x \inr^d
}
and  $(\Sigma^{-1})_{BB} $ is positive definite, it follows easily that $\tilb$ with $\tilb_I=\vk b_I$ and $\tilb_{I^c}=
 \Sigma_{{I^c}I} (\Sigma_{II})^{-1} \b_I$ is the unique solution of $\Pi_\Sigma(\vk b)$,  hence the claim is complete.
\QED

The next result   follows from \cite{MR2397662}[Thm 3.3]  since Gaussian random vectors are particular instances of elliptically symmetric ones where the radius has distribution function in the Gumbel max-domain of attraction with scaling function $w(u)=u$. We present however a short proof. 
\BEL \label{thmL} Let $\vk c\inr^d$ have at least one positive component and let $\vk X$ be a centered $d$-dimensional Gaussian random vector with non-singular covariance matrix $\Sigma$.  Denote by $I,L$ the index sets related to    $\Pi_\Sigma( {\vk c})$ and  let 
further  $ \vk x (u), u> 0$ be a $d$-dimensional vector such that $\limit{u} u^{-1} \vk x(u)= \vk 0.$\\
 As $u\to \IF$ we have 
\bqn{ \label{sh}
	\pk{
		\X_I > (\vk c  u+ \vk x (u))_I} \sim
	\frac{1}{\prod_{i\in I} \vk c_I^\top (\Sigma_{II})^{-1} \vk e_i } u^{- m}
	\varphi_{\vk X_I} ( (\vk c u + \vk x(u) )_I )  , \quad u\to \IF, 
}
where $m$ is the number of elements of $I$ and $\varphi_{\vk X_I}$ is the pdf of $\vk X_I$. Moreover, with 
$N= L \setminus I$  
\bqn{ 
	\limit{u} \frac{ \pk{ \X > \vk c u + \vk x(u)} }{ \pk{ \X_I> (\vk c u+ \vk x(u))_I }} &=&  \limit{u} \frac{ \pk{ \X_L > (\vk c u + \vk x(u))_L} } { \pk{ \X_I> (\vk c u+ \vk x(u))_I } } \notag \\
	&=&
	\pk{ \vk X_N> \vk x_N\ \lvert \vk X_I= \vk x_I}  ,
}  
 provided that $\limit{u} (\vk x(u))_{I\cup N}=\vk x_{I \cup N}$  (set $\pk{ \vk X_N> \vk x_N\ \lvert \vk X_I= \vk x_I} $  to 1 if $N$ is empty).   
\EEL

\BRM In the particular case $\vk x(u)= \vk x/u, \x \inr^d$ from \eqref{sh} we obtain 
\bqny{ 	\pk{\X_I > (\vk c  u+ \vk x /u) _I} \sim 	\pk{\X_I > \vk c_Iu } e^{- \vk x_I^\top  (\Sigma_{II})^{-1}\vk c_I}, \quad u\to \IF.
}
\ERM

\prooflem{thmL}   Assume for simplicity that $I=\{1 \ldot d\}$. In view of \nelem{prop1} $\Sigma^{-1}  \vk c> \vk 0$ and this is the crucial condition for the proof. Note further that $\Pi_{\Sigma}(\vk c)$ has unique solution $\vk c$. Hence for any $u\inr$   we have (set $\vk a(u)= \vk c u + \vk x(u)$)

$$ (\vk a(u) + \vk x/u) ^\top \Sigma^{-1} (\vk a(u)+ \vk x/u) =
( \vk a(u)) ^\top \Sigma^{-1} \vk a(u) + 2 \vk \x ^\top \Sigma^{-1}  \vk a (u)/u + \vk x^\top \Sigma^{-1} \vk x /u^2 .$$
The term $\vk x^\top \Sigma^{-1} \vk x /u^2 $ is important for  showing an integrable upper bound for  the integrand below, and the finiteness of the integral follows from $\Sigma^{-1} \vk c > \vk 0$. More precisely, with $\varphi$ the pdf of $\vk X$  we have 
\bqny{ \pk{ \vk X > \vk a(u)} &=& \int_{ \x > \vk a (u)} \varphi(\x)  d\x \\
	&=& \frac{1}{u^d} \varphi(\vk a(u)) \int_{ \vk y > \vk 0} e^{-  \vk  y ^\top \Sigma^{-1}  \vk a (u)/u + \vk y^\top \Sigma^{-1} \vk y /u^2}d\vk y 
	\sim  \frac{1}{u^d} \varphi (\vk a(u)) \int_{ \vk y > \vk 0} e^{-  \vk  y ^\top \Sigma^{-1}  \vk c} d\vk y
}
since we assume that $\vk x(u)/u \to \vk 0$ as $u\to \IF$. \\
Next suppose that $I$ has $m< d$ elements and let $J=I^c= \{1 \ldot d \} \setminus I$. We have 
\bqny{ \pk{ \vk X > \vk a(u)} &=&  \frac{1}{u^m}\int_{ \vk y_I > \vk 0_I, \vk y_{J} >  (\vk c u - \tilde{\vk c} u)_{J} } \varphi ( \tilde{\vk c} u + \vk x(u) + \vk  y / \bar{\vk u}) d\vk y ,
}
 where $\bar{\vk u}_I=u \vk 1_I$ and $\bar{\vk u}_{J}= \vk 1_{J}$, hence the proof follows easily using further \eqref{coni}. 
 It follows easy that the components of $\vk X$ with indices not in $L$ do not contribute, so we assume without loss of generality that $L$ has $d$ elements. In that case $ (\vk c u - \tilde{\vk c} u)_{J} = \vk{0}_{J}$ and the proof follows after some straightforward calculations.\QED

To this end we prove that $E(\vk c,u)$ in \eqref{27} equals $o(e^{- \varepsilon u^2})$ for some small  $\varepsilon>0$.
We have that 
\bqny{ E(\vk c,u) =o( R(u)), \quad R(u)= \pk{ X_1> c_1 u+ \mu, 	\vk X_{-1} > \vk c_{-1} u} /\pk{ 
			\vk X_{-1} > \vk c_{-1} u } 
		}
		as $u\to \IF$ and $1 \in I$ where the index set $I$ determines the solution of $\Pi_{\Sigma}(\vk c)$. The claim now follows if we show that 
		$\limit{u}R(u)=0$. Indeed this is the case, since in view of \nelem{thmL} the other possibility is that $\limit{u} R(u)= C>0$. 
	This means that the attained minimum of the quadratic programming problem $\Pi_{\Sigma}(\vk c) $ is $\vk c_I ^\top (\Sigma_{II})^{-1}	 \vk c_I$
	being  equal to the attained minimum of $\Pi_{B}(\vk b)$ where $B$ is obtained from $\Sigma$ by deleting the first row and column and $\vk b= \vk c_{-1}$. Since $1 \in I$ there are two different index sets that determine the minimum of the quadratic programming problem $\Pi_\Sigma(\vk c)$ which is a contradiction.

\section*{Acknowledgments}
I am thankful to both referees for detailed review reports which improved the manuscript.  
Support from  SNSF Grant no. 200021-175752/1  is kindly acknowledged.

\bibliographystyle{ieeetr}
\bibliography{EEEA}
\end{document}